\documentclass[review]{elsarticle}

\usepackage[hidelinks]{hyperref}
\usepackage{graphicx}
\usepackage{amsmath}
\usepackage{xcolor}
\newcommand\crule[3][black]{\textcolor{#1}{\rule{#2}{#3}}}
\usepackage{rotating}
\usepackage{float}
\usepackage{numprint}
\usepackage{natbib}
\usepackage{multirow}
\setcitestyle{open={(},authoryear,close={)},aysep={,},citesep={,}}
\usepackage{lineno,hyperref}
\modulolinenumbers[5]

\journal{Elsevier}

\begin{document}

\begin{frontmatter}

\title{Data fusion of distance sampling and capture-recapture data\tnoteref{mytitlenote}}

\author[1]{Narmadha M. Mohankumar\corref{cor1}}
\cortext[cor1]{Corresponding author}
\ead{meenu@ksu.edu}

\author[1]{Trevor J. Hefley}
\author[2]{Katy Silber}
\author[2]{W. Alice Boyle}

\address[1]{Department of Statistics, Kansas State University, 1116 Mid-Campus Drive North, 
Manhattan, Kansas 66506, U.S.A.}
\address[2]{Division of Biology, Kansas State University, 116 Ackert Hall,
Manhattan, Kansas 66506, U.S.A.}

\begin{abstract}

Species distribution models (SDMs) are increasingly used in ecology, biogeography, and wildlife management to learn about the species-habitat relationships and abundance across space and time. Distance sampling (DS) and capture-recapture (CR) are two widely collected data types to learn about species-habitat relationships and abundance; still, they are seldomly used in SDMs due to the lack of spatial coverage. However, data fusion of the two data sources can increase spatial coverage, which can reduce parameter uncertainty and make predictions more accurate, and therefore, can be used for species distribution modeling. We developed a model-based approach for data fusion of DS and CR data. Our modeling approach accounts for two common missing data issues: 1) missing individuals that are missing not at random (MNAR) and 2) partially missing location information. Using a simulation experiment, we evaluated the performance of our modeling approach and compared it to existing approaches that use ad-hoc methods to account for missing data issues. Our results show that our approach provides unbiased parameter estimates with increased efficiency compared to the existing approaches. We demonstrated our approach using data collected for Grasshopper Sparrows (\textit {Ammodramus savannarum}) in north-eastern Kansas, USA. 

\end{abstract}


\end{frontmatter}


\section{Introduction}

Species distribution models (SDMs) are widely used in ecology, biogeography, and wildlife management to learn about species-habitat relationships and estimate abundance across geographic space and time. Inference and predictions from SDMs are increasingly used to inform conservation and management \citep{araujo2006five, kery2015applied, hefley2016hierarchical, koshkina2017integrated}. For example, conflicts between sustaining human activities and preserving biological diversity can be understood by identifying species-habitat relationships across space and time \citep{hefley2015use}. The SDMs are fitted to geo-referenced observations on species such as presence-only, presence-absence, count, distance sampling, and capture-recapture data. Spatially referenced covariates such as elevation, rainfall, soil properties, and vegetation characteristics are used in SDMs to enable statistical inference on species-habitat relationships and obtain spatially heterogeneous abundance estimates \citep{ kery2015applied}.

Distance sampling (DS) and capture-recapture (CR) are two classic types of planned surveys that collect geo-referenced observations on species. The DS data are collected by recording distances to an individual in the study area from a point or transect \citep{burnham1980estimation,burnham1984need,buckland2001introduction}. The CR data are collected by capturing an individual in the study area, which involves physically capturing the individual using a trap (e.g., mist nets) or taking a picture \citep[e.g., camera traps;][]{otis1978statistical,seber1982estimation,pollock1990statistical}. The CR data often contain individual identification where DS data do not. There is a long history of collecting these two types of high-quality planned survey data in the field of ecology and wildlife management. However, DS and CR data are seldomly used in SDMs due to the large amount of effort and cost required to collect data that densely covers a large study area \citep{mcshea2016volunteer}. These two data sources alone may suffer from the lack of spatial coverage, but fusion of the two data sources can increase spatial coverage, which can reduce parameter uncertainty and provide more accurate predictions. Therefore, a fused SDM of DS and CR can provide useful statistical inference regarding the species distribution and abundance more than using any of the data sources alone \citep[see section 25.1 in][]{hooten2019bringing}.

Construction of an adequate fused data SDM for DS and CR data relies upon accounting for missing data issues that are unique to each source of data. Failure to properly account for these missing data issues may lead to misleading inferences and predictions from the SDMs \citep{little1992regression,kery2011towards,dorazio2012predicting,hefley2013nondetection}. The statistical theory and tools to account for missing data issues are well developed in missing data literature, which can be applied to SDMs \citep{ rubin1976inference,little1992regression, mason2010strategy,little2019statistical}, but such tools are rarely explicitly employed in SDM literature \citep{hefley2013nondetection}. Therefore, practitioners often use ad-hoc approaches to account for missing data issues and fit SDMs, which will adversely affect inferences and predictions. In some cases, the ad-hoc techniques may produce biased parameter estimates that invert the inferred species-habitat relationship, which is a critical consequence when making inferences \citep{hefley2014correction, hefley2017bias}. 

Two of the common missing data issues in DS, and CR data are individuals that are missing not at random (MNAR) \citep{little2019statistical} and partially missing location information. The MNAR individuals can occur because of two reasons: 1) limited spatial coverage due to the required large amount of effort and cost, limited accessibility, researcher preferences, or previous knowledge regarding the individual locations, or 2) the individuals in a sampled geographic region being unobserved due to the distance to the individual from the point, transect or the trap, observer's experience level, or environmental or geographical features. The partially missing location information occurs when DS and CR only record partial location information of individuals in contrast to complete location information (e.g., the exact geographic coordinates of the location of the individual). Such partially recorded location information makes spatial covariates unrecoverable because the spatial covariate values are usually obtained from a geographic information system that requires the individuals' exact locations. For example, DS surveys only record the distance to an individual from a point or transect and do not record the exact location of the individual. As another example, CR surveys often use tools to attract the individual to the trap, which results in the original, natural location of the individual being unrecoverable because only the location of the trap is recorded \citep{gerber2012evaluating, williams2018patterns}. Therefore, the spatial covariate values that may influence the locations of the individuals cannot be obtained. 

The missing individuals that are MNAR is implicitly addressed by many DS and CR developments using thinned point process models \citep[e.g.,][]{johnson2010model,borchers2015unifying,fletcher2019practical,farr2020integrating}. Many of these developments use an inhomogeneous Poisson point process (IPPP) which can accommodate spatial inhomogeneity \citep{ diggle1976statistical, cressie2015statistics, kery2015applied} and enable inferences on the species-habitat relationship and abundance \citep{warton2010poisson,renner2015point,hefley2016hierarchical}. The use of the IPPP enables the estimation of an inhomogeneous intensity function that can produce spatial maps showing the species distribution across the study area. In fact, the field of study of SDMs is almost entirely focused on building, fitting, and using models that are capable of estimating an inhomogeneous intensity function to estimate the species distribution \citep[e.g.,][]{warton2010poisson,renner2015point,hefley2016hierarchical}. The spatial maps produced from estimating the species distributions are an essential tool used in conservation reserve planning and administrative regulation implementation \citep[e.g.,][] {hefley2015use}. However, the crux in applying existing IPPP based approaches for DS and CR data is that they may not explicitly address the missing data issues in DS and CR data. For example, the approaches may require complete location information regarding the individuals; however, DS and CR data often contain only partial location information. In practice, researchers use ad-hoc methods to circumvent the limitation of partially recorded locations of individuals and fit the models. For example, \cite{fletcher2019practical} transformed the DS data to presence-absence data at sites using change of support and fitted a model to the presence-absence data. For another example, \cite{farr2020integrating} treated DS data as count data at sampling sites and fitted the model to count data. Both of these approaches do not require complete location information, and the partial location information does not pose an issue since the models are fitted to the transformed DS data. As another example, \cite{borchers2015unifying} proposed an IPPP based unified model for DS and CR data; however, they used a homogeneous point process in all of their applications and did not implement the model for the inhomogeneous case. The homogeneous case contains a constant intensity function where the partial location information is not an issue, but the model is not designed to model the species distribution, which is our primary interest. The inhomogeneous case can model the species distribution; however, the intensity function typically depends on spatially referenced covariates, where the complete location information of the individuals is critical. Therefore, partially recorded location information becomes an issue. In contrast to the ad-hoc approaches, \cite{hefley2020accounting} proposed a model-based approach to account for the partial location information in DS data. However, their model is merely constructed for DS data, and a subsequent model that accounts for the partial location information in CR data is lacking.

In contrast to properly accounting for missing data issues, constructing a fused data SDM requires adequate model representations for DS and CR data that facilitates data fusion. A fused data SDM utilizes information from both types of data to reduce the uncertainty associated with limitations in individual data sources, hence improving the model predictions and inferences \citep{dorazio2014accounting, fithian2015bias, koshkina2017integrated, fletcher2019practical, hooten2019bringing,farr2020integrating}. However, existing IPPP based modeling approaches do not provide model representations for DS and CR data that can be adequately used for data fusion. For example, the unified model proposed by \cite{borchers2015unifying} represented the model for DS data based on the locations of the individuals and represented the model for CR data based on home range centers which are hypothetical centroids for individuals' activity. The locations of home range centers in CR data are irreconcilable with the locations of the individuals in DS data. For example, the model fitted for CR data would estimate the intensity of home range centers, and the model fitted for DS data would estimate the intensity of the locations of the individuals. Therefore, building a fused data SDM where both data sources share parameters in the underlying IPPP targetting the same inference is not achievable. 

A second main issue with existing IPPP based SDMs that involve data fusion is that they often perform spatial aggregation. Spatial aggregation involves partitioning the study area and transforming the locations of the individuals to counts in each of the partitions \citep[e.g.,][]{dorazio2014accounting, koshkina2017integrated,farr2020integrating}. However, a significant drawback of spatial aggregation is determining the spatial resolution for the partitions. If the spatial resolution does not adequately represent the sampled region, the model may yield biased estimates of parameters and abundance. 

We propose a hierarchical modeling framework that provides adequate model representations for DS and CR data enabling data fusion and targeting the equivalent inference regarding species-habitat relationship and abundance. We use theory and tools from the missing data literature to build models for the missing data mechanism and account for the missing data issues. Our modeling framework can be viewed as a unified framework that can be applied to many other data sources (e.g., presence-only data) and a fusion of them addressing critical issues with missing data. Therefore, our approach advances the types of models developed for species distribution studies. In our work, we propose two fused data SDMs for DS and CR data, one SDM incorporating the recorded distances from DS data and the other SDM without incorporating the recorded distances. We compare the two SDMs and investigate the efficiency gain of the estimated parameters by incorporating additional information regarding the observed individuals, such as the recorded distances. We conduct a simulation experiment to evaluate the performance of our two SDMs compared to existing approaches that use spatial aggregation. We assess the accuracy and the efficiency of the estimated parameters for the species-habitat relationship and obtain an estimate for the expected abundance in the study area. Finally, we demonstrate the approaches using data collected for Grasshopper Sparrows (\textit {Ammodramus savannarum}) in North-Eastern Kansas.

\section{Materials and Methods}

\subsection{Hierarchical modeling framework}

Our proposed fused data SDM relies on a hierarchical modeling framework that is based on an IPPP. The models for the observed DS and CR data are conditioned on a common underlying IPPP that represents the underlying point pattern of individuals in the study area. 

\subsubsection{The underlying IPPP}

The underlying IPPP describes the random number and the locations of individuals across the study area based on a continuous inhomogeneous intensity function, a function of spatially referenced covariates (e.g., elevation, temperature, soil attributes, vegetation, etc.). The intensity describes the expected number of individuals per infinitely small unit area and is usually defined as $\lambda(\mathbf s) =  e^{\mathbf{x}(\mathbf {s})'\boldsymbol \beta}$, where, $\textbf s$ represents a vector containing coordinates of a location within the study area $\mathcal{S}$,  $\mathbf
{x}(\textbf s) \equiv (1, x_1(\textbf s), x_2(\textbf s), ..., x_q(\textbf s))'$, and $\boldsymbol{\beta} \equiv (\beta_0, \beta_1, \beta_2, ..., \beta_q)'$. The $x_1(\textbf s), x_2(\textbf s), ..., x_q(\textbf s)$ represent the spatial covariates at the location $\textbf s$, $\beta_0$ represents the intercept parameter, and $ \beta_1, \beta_2,... , \beta_q$ represent the regression coefficients associated with the species-habitat relationship. Using the above notation, the probability distribution function (PDF) for the IPPP can be written as \citep{cressie2015statistics}

\begin{equation}
[\textbf{u}_1, \textbf{u}_2, ..., \textbf{u}_N, N|\lambda(\mathbf s)] =  \frac{e^{-{\int_{\mathcal{S}}\lambda(\mathbf s)d \mathbf s}}(\int_{\mathcal{S}}\lambda(\mathbf s)d \mathbf s)^N}{N!} \times 
N! \prod_{i=1}^{N} \frac{\lambda(\textbf u_i)}{\int_{\mathcal{S}}\lambda(\mathbf s)d \mathbf s},
\end{equation}

\noindent where  $\textbf{u}_1, \textbf{u}_2, ..., \textbf{u}_N$ are the locations of all $N$ individuals (missing and observed) in the study area $\mathcal{S}$ (i.e.,  $\textbf{u}_i \in \mathcal{S}$). A property of IPPP is that an estimate of the expected abundance in any sub-region $\mathcal{B}$ in the study area can be represented by $\bar\lambda = \int_{\mathcal{B}} e^{\mathbf{x}(\mathbf {s})'\boldsymbol \beta}d \mathbf s$.

\subsubsection{Accounting for missing individuals that are MNAR}

The missing individuals that are MNAR can be accounted for by identifying and modeling the missing data mechanism. To model the missing data mechanism, we can label the random locations of all individuals in the study area as missing or observed \citep{gelfand2018bayesian}. We can define a vector $\textbf m = (m(\textbf{u}_1), m(\textbf{u}_2), ... m(\textbf{u}_N))$, where $m(\textbf{u}_i)$ labels the $i^
\text{th}$ individual as missing (i.e., zero) or observed (i.e., one). Employing the missing data mechanism, we can write the distribution of $m(\textbf{u}_i)$ as a zero-inflated Bernoulli distribution conditioned on $\textbf{u}_i$.

\begin{equation}
[m(\textbf{u}_i)|\textbf{u}_i,  q(\mathbf s),  r(\mathbf s)] = \left \{\begin{matrix}
 q(\mathbf{u}_{i})^{m(\textbf{u}_i)}(1-q(\mathbf{u}_{i}))^{1-m(\textbf{u}_i)}  & ,\text{if}  \; r(\textbf{u}_i) = 1
\\ 
0
& ,\text{if}  \; r(\textbf{u}_i) = 0
\end{matrix}\right.,
\end{equation}

\noindent where, $q(\mathbf{u_i})$ denote the probability of observing the individual, $r(\textbf{u}_i) =1$ denotes that the $\textbf{u}_i^\text{th}$ location is sampled within the study area, and $r(\textbf{u}_i) =0$ denotes that the $\textbf{u}_i^\text{th}$ location is not sampled within the study area. The functional form of $q(\mathbf s)$ and $r(\mathbf s)$ at a location $\mathbf s$ can be defined based on the missing mechanism. The $r(\mathbf s)$ accounts for the missing individuals that are MNAR due to unsampled geographic regions in the study area, and $q(\mathbf s)$ accounts for the missing individuals that are MNAR when the corresponding geographic region is sampled, but the individuals are not detected or captured.

By using the distribution of $m(\textbf{u}_i)$, we can derive the PDF for the location of the  $i^
\text{th}$ individual conditioned on the label $m(\textbf{u}_i)$ as

\begin{equation}
\begin{aligned}
{}[\textbf{u}_i| m(\textbf{u}_i), \lambda(\mathbf s), q(\mathbf s),  r(\mathbf s)] =
\left \{\begin{matrix}
 \frac{q(\mathbf{u}_{i})^{m(\textbf{u}_i)}(1-q(\mathbf{u}_{i}))^{1-m(\textbf{u}_i)}  \lambda(\textbf u_i)} {\int_{\mathcal{S}}q(\mathbf s)^{m(\textbf s)}(1-q(\mathbf s))^{1-m(\textbf s)}   \lambda(\mathbf s)d \mathbf s} & ,\text{if} \; r(\textbf{u}_i) = 1
\\ 
0
& ,\text{if} \; r(\textbf{u}_i) = 0
\end{matrix}\right..
\end{aligned}
\end{equation}

\noindent An important property of the distributional representation in (3) is that it enables the estimation of the locations of unobserved individuals in addition to the locations of the observed individuals. The locations of unobserved individuals can be estimated by augmenting the unobserved individuals and modeling using a Bayesian framework. Many recent model-based approaches based on IPPP use the so-called thinned IPPP \citep{diggle1976statistical,chakraborty2011point,cressie2015statistics, kery2015applied}, an implicit representation of the data to account for missing individuals as opposed to the complete distributional representation in (3). 

\subsubsection{Accounting for partially missing location information}

The distributional representation in (3) accounts for the missing individuals that are MNAR; however, it does not account for the partially missing location information. It requires complete location information of the individuals. We propose two models to account for the partially observed location information in data; 1) a model without incorporating the recorded distances from DS, and 2) a model incorporating the recorded distances from DS.

The DS and CR surveys each contain a sampled region in the study area, a region surrounding the points, transects, or the traps where the probability of detection or capture is greater than zero. We denote this region as the detection/capture region. In our first proposed model, we assume that the observed location of an individual is uniformly distributed in the detection/capture region that surrounds the point, transect, or the trap it was observed. Under this assumption, we can write the PDF of the observed location of the $i^\text{th}$ individual conditioned on the actual location of the individual as

\begin{equation}
 [\textbf{y}_{i}|\textbf{u}_i] = \left \{\begin{matrix}
|{A}_{u_i}|^{-1} I (\mathbf{y}_i\in A_{u_i}) &, \text{if} \;  m(\textbf{u}_i)=1
\\ 
0
&, \text{if} \;  m(\textbf{u}_i)=0
\end{matrix}\right.,
\end{equation}
where, $\textbf y_{i}$ denote the observed location of the $i^{\text{th}}$ individual, $\textbf u_{i}$ is the actual location of the  $i^{\text{th}}$ individual, and $A_{u_i}$ is the detection/capture region surrounding the point, transect or the trap where the individual was observed.

We then propose a second model by incorporating the recorded distances from DS data into the model. We expect that adding additional information regarding the observed locations of the individuals may increase the efficiency of the model parameter estimates. \cite{hefley2020accounting} account for the partial location information in DS data by incorporating the recorded distances. Based on their approach, and under the assumption that the distances are recorded perfectly, we can assume that the observed location of an individual from a transect is uniformly distributed along the parallel lines to the transect ($L_{u_i}$) with a perpendicular distance that is equal to the recorded distance $d_i$. Under this assumption, we can write the PDF of the observed location of the $i^\text{th}$ individual conditioned on the actual location of the individual as

\begin{equation}
 [\textbf{y}_{i}|\textbf{u}_i] = \left \{\begin{matrix}
|{L}_{u_i}|^{-1} I (\mathbf{y}_i\in L_{u_i}) &,\text{if} \; m(\textbf{u}_i)=1
\\ 
0
&,\text{if} \; m(\textbf{u}_i)=0
\end{matrix}\right..
\end{equation}
 
\noindent For a point, $L_{u_i}$ is the perimeter of the circle, where the radius is equal to the recorded distance, $d_i$. The $|L_{u_i}|$ is the length of the lines or the length of the perimeter of the circle.

\subsection{Model implementation}

The distributions in (4) and (5) represent the observed location of the $i^\text{th}$ individual conditioned on the actual location of the observed individual, $\textbf u_{i}$; however, the actual location of the observed individual is of little interest in our study. Therefore, we can remove $\textbf u_{i}$ from the model by integrating the joint likelihood of $\mathbf{y}_{i}$ and $ \textbf u_i$. The resulting PDFs representing the observed location of the $i^\text{th}$ individual are

\begin{equation}
 [\textbf{y}_{i}|m(\textbf{u}_i),\lambda(\mathbf s),q(\mathbf s),r(\mathbf s)] = \left \{\begin{matrix}
\frac{\int_{A_{u_i}} |{A}_{u_i}|^{-1}\lambda(\textbf u_{i})q(\mathbf{u}_{i})d\mathbf{u}_{i}}{\int_{\mathcal{S}} \lambda(\textbf s)q(\mathbf{s})d\mathbf{s}} &, \text{if} \;  r(\textbf{u}_i)=1 \;  \& \; m(\textbf{u}_i)=1
\\ 
0
& , otherwise
\end{matrix}\right., 
\end{equation}

\begin{equation}
 [\textbf{y}_{i}|m(\textbf{u}_i), \lambda(\mathbf s),q(\mathbf s),r(\mathbf s)] = \left \{\begin{matrix}
\frac{\int_{L_{u_i}} |{L}_{u_i}|^{-1}\lambda(\textbf u_{i})q(\mathbf{u}_{i})d\mathbf{u}_{i}}{\int_{\mathcal{S}} \lambda(\textbf s)q(\mathbf{s})d\mathbf{s}} &, \text{if} \;   r(\textbf{u}_i)=1 \;  \& \; m(\textbf{u}_i)=1
\\ 
0
&, otherwise
\end{matrix}\right..
\end{equation}

Moreover, our objectives in the study do not focus on estimating the locations of the unobserved individuals. Therefore, we can retain the PDF for the observed individual locations from (6) and (7) by setting $m(\textbf{u}_i))=1$. The resulting PDF is a simple marginal distribution that can be fitted using a likelihood-based or Bayesian approach. If practitioners are interested in estimating the locations of unobserved individuals, they can fit the model using a Bayesian hierarchical modeling approach from (3–5). Details associated with deriving our models are provided in the Supplementary Material.

\subsection{Fused data SDM}

The distributional representations in (6) and (7) can be used to construct a fused data SDM for DS and CR data. Our proposed distributional represntations represent both DS and CR data based on observed locations of the individuals; therefore, the models share parameters in the underlying IPPP that target the same inference. We assume that the observed locations in the DS and CR data are independent across points, transects and traps within and between the surveys. Representing DS and CR data using our proposed distributional representations and jointly modeling them leads to the following two fused data SDMs. The distribution in (8) does not incorporate the recorded distances from DS data, and the distribution in (9) incorporates the recorded distances.

\begin{equation}
\begin{aligned}
{}[\textbf{y}_{1}, ..., \textbf{y}_{n_{ds}}, \textbf{y}_{n_{ds}+1}, ..., \textbf{y}_{n_{ds}+n_{cr}}, n_{ds}, n_{cr}|\lambda(\mathbf s),q_{ds}(\mathbf s),r_{ds}(\mathbf s), q_{cr}(\mathbf s),r_{cr}(\mathbf s)] =\\
e^{-\int_{\mathcal{S}}\lambda(\mathbf s)q_{ds}(\mathbf s)I(r_{ds}(\mathbf s)=1)d \mathbf s -\int_{\mathcal{S}}\lambda(\mathbf s)q_{cr}(\mathbf s)I(r_{cr}(\mathbf s)=1)d \mathbf s}  \times \\
\prod_{i=1}^{n_{ds}}  \int_{A_{u_i}} |{A}_{u_i}|^{-1}\lambda(\textbf u_{i})q_{ds} (\mathbf{u}_{i})I(r_{ds}(\textbf{u}_i)=1) \times \\
\prod_{i=n_{ds}+1}^{n_{ds}+n_{cr}}  \int_{A_{u_i}} |{A}_{u_i}|^{-1}\lambda(\textbf u_{i})q_{cr}(\mathbf{u}_{i})I(r_{cr}(\textbf{u}_i)=1),
\end{aligned}
\end{equation}

\begin{equation}
\begin{aligned}
{}[\textbf{y}_{1}, ..., \textbf{y}_{n_{ds}}, \textbf{y}_{n_{ds}+1}, ..., \textbf{y}_{n_{ds}+n_{cr}}, n_{ds}, n_{cr}|\lambda(\mathbf s),q_{ds}(\mathbf s),r_{ds}(\mathbf s), q_{cr}(\mathbf s),r_{cr}(\mathbf s)] =\\
e^{-\int_{\mathcal{S}}\lambda(\mathbf s)q_{ds}(\mathbf s)I(r_{ds}(\mathbf s)=1)d \mathbf s -\int_{\mathcal{S}}\lambda(\mathbf s)q_{cr}(\mathbf s)I(r_{cr}(\mathbf s)=1)d \mathbf s}   \times \\
\prod_{i=1}^{n_{ds}}  \int_{L_{u_i}} |{L}_{u_i}|^{-1}\lambda(\textbf u_{i})q_{ds} (\mathbf{u}_{i})I(r_{ds}(\textbf{u}_i)=1) \times \\
\prod_{i=n_{ds}+1}^{n_{ds}+n_{cr}}  \int_{A_{u_i}} |{A}_{u_i}|^{-1}\lambda(\textbf u_{i})q_{cr}(\mathbf{u}_{i})I(r_{cr}(\textbf{u}_i)=1),
\end{aligned}
\end{equation}

\noindent  where, $n_{ds}$ and $n_{cr}$ are the number of detected and captured individuals from DS and CR respectively,  $q_{ds}(\cdot)$ is the probability of detection from a point or transect which depends on the distance from the point or transect to the individual,  $q_{cr}(\cdot)$  is the probability of capture from a trap, $r_{ds}(\text s)$ and $r_{cr}(\text s)$ are indicator functions defining the detection/capture regions of the DS and CR data respectively, and $n= n_{ds} + n_{cr}$ is the total number of observed individuals from surveys. In our study, we define the probability of detection for DS data by a half-normal function, that is $q_{ds}(\mathbf{u}_{i}) =  e^{- d_{i}^2/\phi}$, where, $d_{i}$ is the distance between the point or transect and the the $i^\text {th}$ detected individual, and $\phi$ is a scale parameter. The indicator function truncating the detection region from a point or transect is defined as, $ r_{ds}(\mathbf{u}_{i}) = I (\mathbf{u_i}\in A_{ds}),
$ where $A_{ds}$ is the detection region surrounding a point or transect where probability of detection is greater than zero. We define the probability of capture from a trap as $q_{cr}(\mathbf{u}_{i})= \theta$. The indicator function truncating the capture region of a trap is defined as $r_{cr}(\mathbf{u}_{i})= I (\mathbf{u}_i\in A_{cr})$, where $A_{cr}$ is the capture region surrounding a trap where probability of capture is greater than zero.

In principle, including additional information regarding the observed individual locations ought to increase the efficiency of parameter estimates from a model. Therefore, we expect the fused data SDM in (9) to provide more efficient parameter estimates than the fused data SDM (8) since the SDM in (9) incorporates the recorded distances from DS data. We investigate this fact in both the simulation experiment and the data example that follows. 

\section{Simulation experiment}

We conducted a simulation experiment to evaluate the performance of our two proposed fused data SDMs and compare them to standard approaches that use spatial aggregation. We assessed the performance of the models using the five scenarios listed below.

\begin{enumerate}
  \item The model from (3) fit to DS and CR data containing complete location information of the individuals.
  \item The model proposed by \cite{farr2020integrating} for spatially aggregated data fit to DS and CR data containing partial location information of the individuals.
  \item The model from (3) tranformed for spatially aggregated data using change of support fit to DS and CR data containing partial location information of the individuals.
  \item Our proposed fused data SDM from (8) without incorporating recorded distances fit to DS and CR data containing partial location information of the individuals.
  \item Our proposed fused data SDM from (9) incorporating recorded distances fit to DS and CR data containing partial location information of the individuals.
\end{enumerate}

In our simulation experiment, we simulated a single spatial covariate, $x(\textbf{s}$) using a reduced rank Gaussian process on an unit square study area (i.e., $\mathcal{S} =[0, 1] \times [0, 1]$, where $\textbf s \in \mathcal{S})$. We simulated the actual locations of the individuals using the IPPP represented by (1) with the intensity $\lambda(\mathbf s) = e^{\mathbf{x}(\mathbf {s})'\boldsymbol \beta}$. We set the parameter values as $\beta_0 = 9, \beta_1 =1, \theta=0.2,$ $\phi=0.025$. We placed 15 points and 65 traps in the study area to obtain DS and CR data, respectively (Fig.1; panel a). We set non-overlapping detection/capture regions to ensure the independence of the observed data across surveys and within surveys (Fig.1; panel c). We constructed the detection region surrounding each point by defining that the individual has to be within a maximum distance of 0.04 from the point to be detected. We constructed the capture region surrounding each trap by defining that the individual has to be within a maximum distance of 0.02 from the trap to be attracted and captured. We obtained spatially aggregated data required to fit the models in scenario 2 and scenario 3 by dividing the study area into 100 non-overlapping partitions and obtaining the number of observed individuals in each partition (Fig.1; panel b). If a partition does not consist of a survey point or a trap, we defined the partition as an unsampled partition.

We simulated 1000 data sets and fitted the models described in scenarios 1–5. We used the complete location information of the individuals in scenario 1, whereas the partial location information of the individuals in scenarios 2–5. Scenario 1 acts as the benchmark scenario where the data with complete location information matches the process described by the fitted model. We evaluated the performance of the models in scenarios 2–5 for data containing partial location information and compared them to benchmark scenario 1. For each simulated data set, we obtained the parameter estimates for the intercept ($\beta_0$), the relationship to the spatial covariate ($\beta_1$), and the expected abundance (${\bar\lambda}$). We assessed the reliability of the parameter estimates by calculating the coverage probabilities of the 95\% Wald-type confidence intervals (CIs). We included side-by-side box plots to visually compare the empirical distributions of the parameter estimates. We obtained the relative efficiency of the parameter estimates under scenarios 2–5 with reference to the efficiency of parameter estimates obtained under benchmark scenario. The relative efficiency is calculated by dividing the standard deviation of the respective empirical distribution of the estimates by the standard deviation of the empirical distribution of the estimates under scenario 1.

The integrals in the likelihood functions and the integrated intensity function are approximated using numerical quadrature. We used the Nelder-Mead algorithm in R to numerically maximize the likelihoods and obtain the parameter estimates $\hat\beta_0$ and $\hat\beta_1$. The estimate for the expected abundance is obtained using $ \hat {\bar\lambda} =\int_{\mathcal{S}}e^{\mathbf{x}(\mathbf {s})'\hat {\boldsymbol \beta}}d\mathbf {s} $. We inverted the Hessian matrix to approximate the standard errors of the parameter estimates $\hat\beta_0$ and $\hat\beta_1$ and then calculated the 95\% Wald-type CIs for $\hat\beta_0$ and $\hat\beta_1$. We approximated the standard error of the parameter estimate $ \hat {\bar\lambda}$ using the delta method under first-order Taylor expansion and then calculated 95\% Wald-type CI for $ \hat {\bar\lambda}$. We provide the annotated R code associated with the simulation experiment in the Simulation.R file in the Supplementary Material.

\begin{figure}
    \centering
    \includegraphics[width=\textwidth, clip]{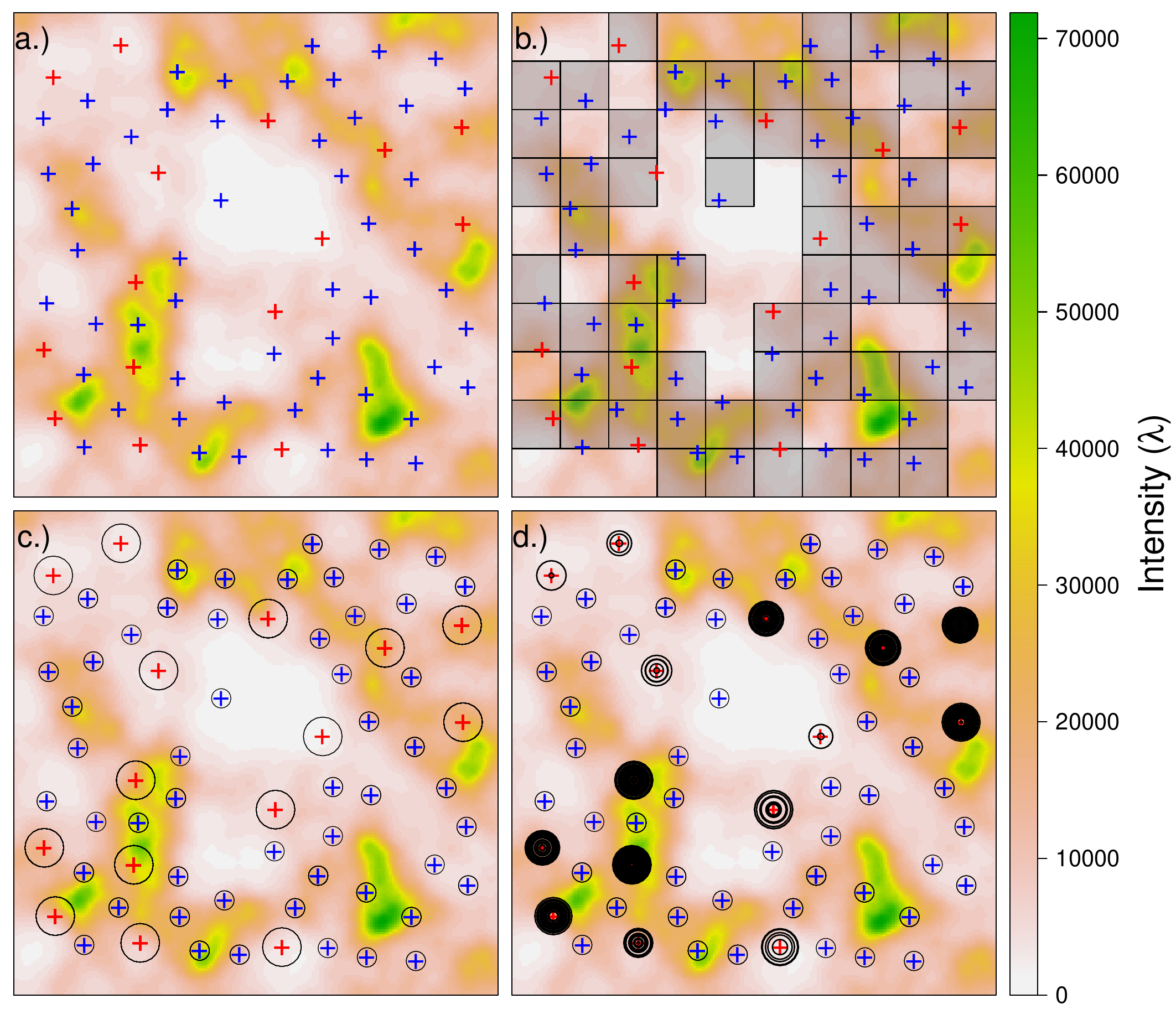}
    \renewcommand\thefigure{1} 
   \caption{Panel (a) displays the points (red $\color{red}+$) and traps (blue $\color{blue}+$) placed in the study area to collect DS and CR data. Panel (b) shows the partitioning of the study area to obtain spatially aggregated DS and CR data (for scenario 2 and scenario 3). Spatially aggregated data are obtained by dividing the study area into 100 non-overlapping partitions and choosing the partitions that include a point or a trap. Panel (c) displays the detection and capture regions of DS and CR data (for scenario 4). Panel (d) displays the circle's perimeter surrounding the points, where the radius is equal to each individual's recorded distance (for scenario 5). Panel (d) also displays the capture regions of the traps.}
\end{figure}

\section{Grasshopper Sparrows at Konza Prairie Biological Station, Kansas}

We illustrated our proposed models and the existing approaches using data on Grasshopper Sparrows (\textit {Ammodramus savannarum}) from Konza Prairie Biological Station (KPBS). KPBS is a long-term ecological research site in northeastern Kansas, comprised of native tallgrass prairie \citep{knapp1998grassland, williams2018patterns, williams2019causes}. Grasshopper Sparrows are a migratory grassland songbird species that winter in the Southern United States and Northern Mexico and breed throughout grasslands in the United States and Southern Canada. However, the loss of prairie habitat has contributed to a long-term population decline in Grasshopper Sparrows \citep{herse2018importance}. Therefore, identifying suitable habitats and investigating the abundance of Grasshopper Sparrow populations is essential for directing conservation efforts.

We used observations from the 2019 breeding season for our analysis. The data consist of 72 observations from 53 transects and 160 observations from 137 mist-net locations (Fig. 2; panel a). The transects were surveyed during the month of June as part of the long-term monitoring efforts of birds at the Konza Prairie. Within 24 experimentally-managed pastures, one to four 300m long transects bisect the topographic gradients within the sampling site. A single observer slowly walks the transect, recording the individuals seen or heard on either side of the transect, with the distance to each individual \citep{boyle2019cbp01}. The mist-nets were used to capture individuals during the entire breeding season from shortly after the adult male birds arrive in April until nests complete in August. The mist net locations were selected to maximize chances of capturing the adult male birds within their territories, and the birds were attracted to nets using a small speaker broadcasting a territorial song \citep{williams2018patterns}. 

Male adult birds sing territorial songs from conspicuous perches in suitable habitats and actively defend ~0.5 ha territories from other male birds \citep{winnicki2020social}. Female birds select and build nests within the territories of male birds. Their behavior is very secretive, making them difficult to detect. Thus, both detections and captures consist of male adult birds only. Upon arrival, the male adult birds establish breeding territories at the site. These individual male adult birds may select territories based on many environmental cues such as vegetation, topography, location of conspecifics, and land management \citep{andrews2015use, shaffer2021effects}. To illustrate our approach, we use elevation as the spatial covariate.

We illustrate our approach for DS and CR data using the detections from transects and captures from mist-nets. We assume that the individual has to be within a maximum distance of 150m from the transect to be detected, which is realistic given the topography, song attenuation, and realized distance values (Fig. 2; panel c). For captures from mist-nets, we assume that the individual has to be within a maximum distance of 25m to elicit a response and be attracted to the mist-net, a distance reasonable given the speaker volume and observed behavior of the species (Fig. 2; panel c). Furthermore, we assume that the observations from the transects and the mist-nets are independent within and between the surveys.

As in scenarios 2–5 in the simulation experiment, we fit the four models to the observed data: 1) the model proposed by Farr et al. (2020) for spatially aggregated data, 2) the model from (3) transformed for spatially aggregated data using change of support, 3) our proposed fused data SDM from (8) without incorporating recorded distances, and 4) our proposed fused data SDM from (9) incorporating recorded distances. We obtain the spatially aggregated data by dividing the study area into non-overlapping partitions and counting observed individuals in each partition. If a partition does not consist of a transect or a mist net, we define the partition as an unsampled partition which led to 66 non-overlapping sampled partitions (Fig. 2; panel b). Finally, we fit the models to the data, compare the maximum likelihood estimates and the corresponding 95\% Wald-type CIs for $\beta_0$, $\beta_1$, and $\bar\lambda$. We provide the annotated R code associated with the data analysis in the Grasshopper\_sparrows\_data\_example.R in the Supplementary Material.

\begin{figure}
    \centering
    \includegraphics[width=\textwidth, clip]{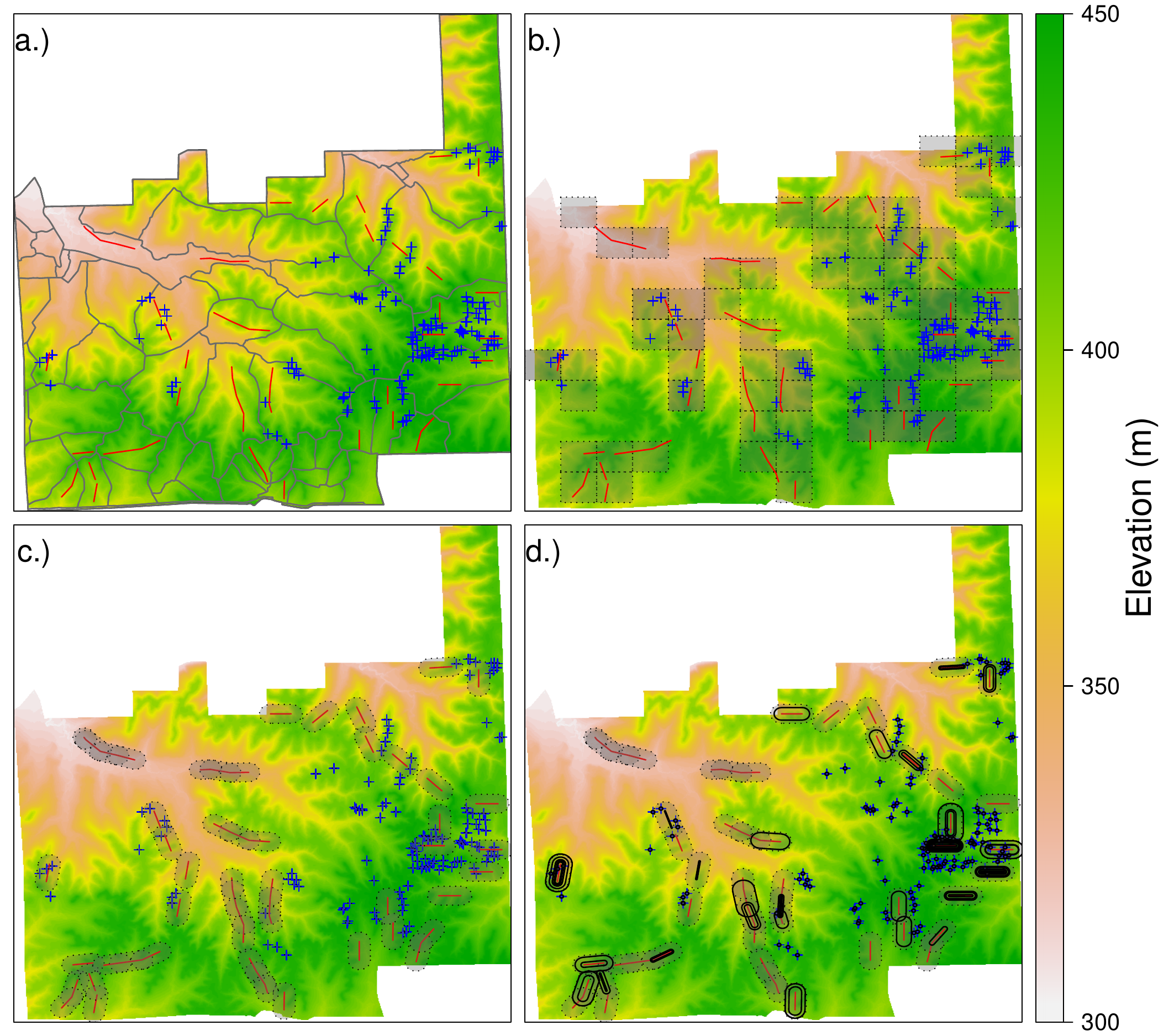}
    \renewcommand\thefigure{2} 
   \caption{Panel (a) displays the transects (red $\color{red}\textendash$) and mist nets (blue $\color{blue}+$) that are used to collect data on Grasshopper Sparrows at Konza Prairie Biological Station (KPBS). The surveys are conducted at watershed-level (grey $\color{gray}\textendash$ in panel (a)). Panel (b) shows the partitioning of the study area (66 partitions) to obtain spatially aggregated data (dashed line) to fit the two models; the model proposed by \cite{farr2020integrating} for spatially aggregated data, and the model from (3) transformed for spatially aggregated data using change of support. Panel (c) displays the detection and capture regions of transects and traps (dashed line) used for our proposed fused data SDM from (8) without incorporating recorded distances. Panel (d) displays the parallel lines to the transect with a perpendicular distance equal to each individual's recorded distance, which is used for our proposed fused data SDM from (9) incorporating recorded distances. Panel (d) also displays the capture regions of the traps (dashed line).
}
\end{figure}

\section{Results}

\subsection{Simulation experiment}

As expected, the benchmark scenario (i.e., scenario 1) yielded an unbiased estimate for $\beta_0$, with a high coverage probability of the 95\% CIs, 0.942. When the data contained partial location information, scenario 2 and scenario 3 yielded biased estimates for $\beta_0$, whereas scenario 4 and scenario 5 yielded unbiased estimates (see Fig. 3 for graphical comparison). The coverage probabilities of the 95\% CIs for $\beta_0$ under scenarios 2–5 were 0.190, 0.180, 0.761, and 0.925, respectively. The relative efficiencies of estimates for $\beta_0$ obtained from scenarios 2–5 were 23.204, 15.949, 13.907, and 1.007, respectively. We noticed that the efficiency of the estimate for $\beta_0$ under scenario 5, surprisingly reaches the efficiency obtained under the benchmark scenario 1 (see Table. 1). 

Similar to the parameter estimate for $\beta_0$, scenario 1 yielded an unbiased estimate for $\beta_1$ with a high coverage probability of the 95\% CIs,  0.948. However, when the data contained partial location information, scenario 2 and scenario 3 yielded biased estimates for $\beta_1$, whereas scenario 4 and scenario 5  yielded unbiased estimates for $\beta_1$ (see Fig. 3 for graphical comparison). The coverage probabilities of the 95\% CIs for $\beta_1$ under scenarios 2–5 were 0.749, 0.838, 0.942 and 0.942, respectively. The relative efficiencies of estimates for $\beta_1$ obtained from scenarios 2–5 were 1.891, 1.394, 1.089, and 1.041, respectively, where scenario 5 provides the most efficient parameter estimate for $\beta_1$ (see Table. 1).

Scenario 1 yielded an unbiased estimate for $\bar\lambda$ with a high coverage probability of the 95\% CIs, 0.944. When the data contained partial location information, scenario 4 and scenario 5 yielded unbiased estimates for $\bar\lambda$. The coverage probabilities of the 95\% CIs for $\bar\lambda$ under scenarios 2–5 were 0.343, 0.430, 0.783, and 0.944, respectively. The relative efficiencies of the estimates for $\bar\lambda$ obtained from scenarios 2–5 were 265.921, 285.819, 141.896, and 1.038, respectively. We noticed that scenario 5 provides the most efficient parameter estimate for $\bar\lambda$, which surprisingly reaches the efficiency obtained under benchmark scenario 1  (see Table. 1). 

\begin{figure}
    \centering
    \includegraphics[width=5in, clip]{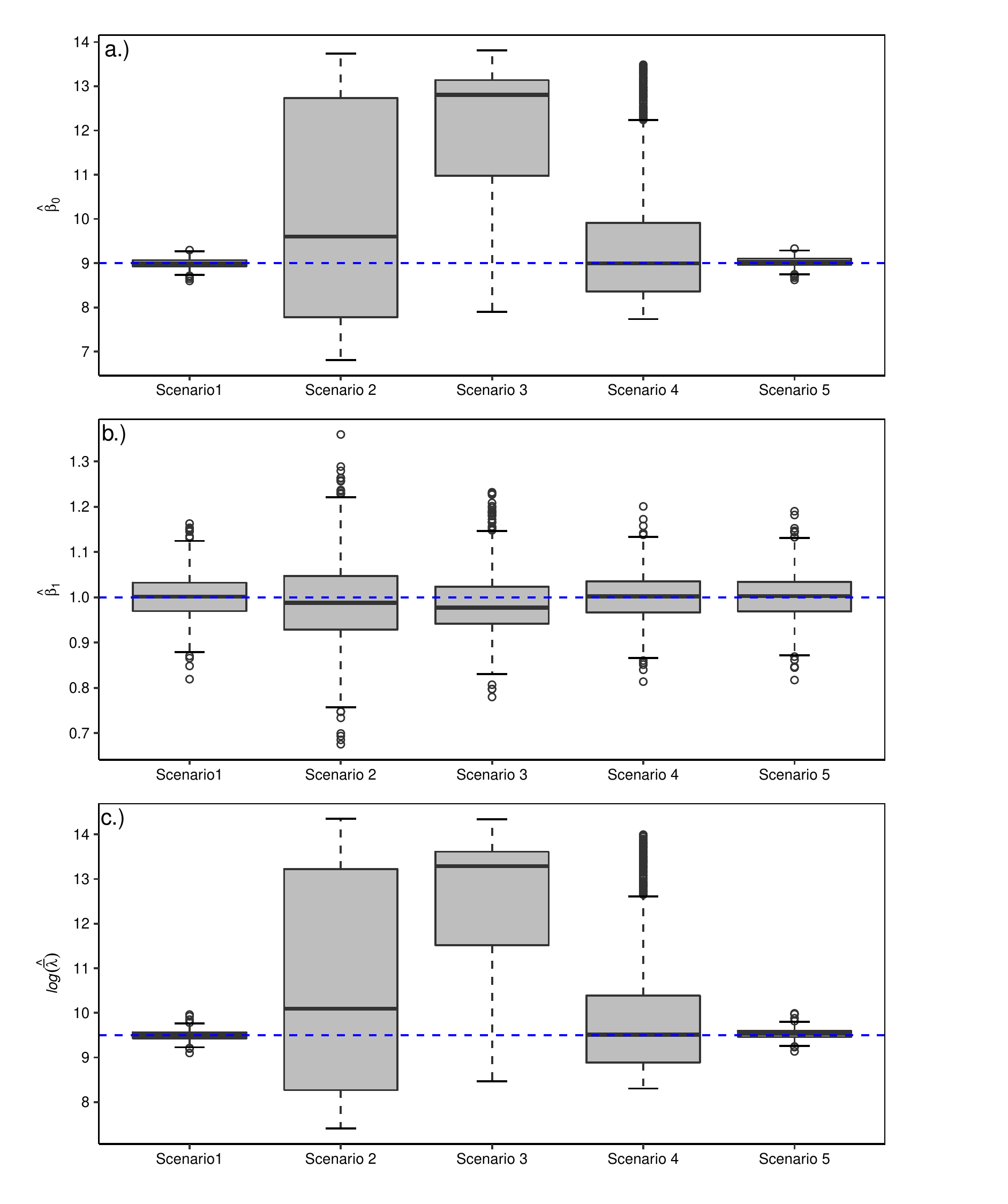}
    \renewcommand\thefigure{3} 
\caption{The box plots display the estimates of parameters $\beta_0$ (panel a), $\beta_1$ (panel b), and $log(\bar\lambda)$ (panel c) obtained under scenarios 1–5 for 1000 simulated data sets. The true values of the parameters ($\beta_0$ = 9, $\beta_1$ =1, $log(\bar\lambda)$= 9.5 ) are shown by the blue dash line ($\color{blue}\textendash$).
}
\end{figure}

\renewcommand{\arraystretch}{1}
\begin{table}[H]
\centering
\renewcommand\thetable{1} 
\caption{Estimated coverage probability (CP) for the 95\% confidence interval (CI) and the relative efficiency (RE) for the parameters $\beta_0$, $\beta_1$, and expected abundance ($ {\bar\lambda}$) obtained under scenario1, scenario 2 , scenario 3, scenario 4, and scenario 5 in the simulation experiment. The parameter estimates are obtained by fitting the models to 1000 simulated data sets.
}
\npdecimalsign{.}
\nprounddigits{2}

\begin{tabular}
{p{0.16\textwidth}|>
{\centering}p{0.1\textwidth}>
{\centering}p{0.1\textwidth}|>
{\centering}p{0.1\textwidth}>
{\centering}p{0.1\textwidth}|>
{\centering}p{0.1\textwidth}>
{\centering\arraybackslash}p{0.1\textwidth}}
\hline
\multirow{2}{*}{Scenarios} &\multicolumn{2}{c}{$\beta_0$}&\multicolumn{2}{c}{$\beta_1$}&\multicolumn{2}{c}{$\bar\lambda$}\\\cline{2-7}
& CP & RE & CP & RE & CP & RE\\
\hline
  Scenario 1 & 0.942 & - & 0.948 & -  &0.944 & -\\
  \hline
    Scenario 2 & 0.190 & 23.204 & 0.749 & 1.891 & 0.343& 265.921\\
    \hline
    Scenario 3 & 0.180 & 15.949 & 0.838 & 1.394& 0.430 & 285.819\\
    \hline
   Scenario 4 & 0.761 & 13.907 & 0.942& 1.089& 0.783 & 141.896\\
   \hline
  Scenario 5 & 0.925 & 1.007 & 0.942 & 1.041 & 0.944 & 1.038 \\
\hline
\end{tabular}
\end{table}

\subsection{Grasshopper Sparrows at Konza Prairie Biological Station,Kansas}

The estimates obtained for the intercept parameter ($\beta_0$) under our two proposed models were similar, with narrow 95\% CIs. The models that use spatially aggregated data yielded similar estimates for $\beta_0$, but with approximately 12 times wider CIs than our proposed models (see Fig. 4; panel a, and 95\% CIs in Table 2). The estimates obtained for $\beta_1$ under all four models yielded similar inference regarding the relationship between species abundance and elevation; however, the estimate for $\beta_1$ under the model proposed by \cite{farr2020integrating} was twice as large as the estimates obtained from the other models (see Fig. 4; panel b, and 95\% CIs in Table 2). The crucial outcome from our fitted models is the estimates obtained for $\bar\lambda$. The models that use spatially aggregated data yielded inexplicable estimates for $\bar\lambda$ with an approximate 163000 times wider 95\% CIs than our proposed models (see Fig. 4; panel c, and 95\% CIs in Table 2). Altogether, the parameter estimates $\hat\beta_0$, $\hat\beta_1$, and $\hat{\bar\lambda}$ from our proposed two models were similar and yielded narrower 95\% CIs. The similarity of the estimates obtained from our two models may be due to the smooth surface of the spatial covariate "elevation."

\begin{sidewaysfigure}
\includegraphics[width=8.in]{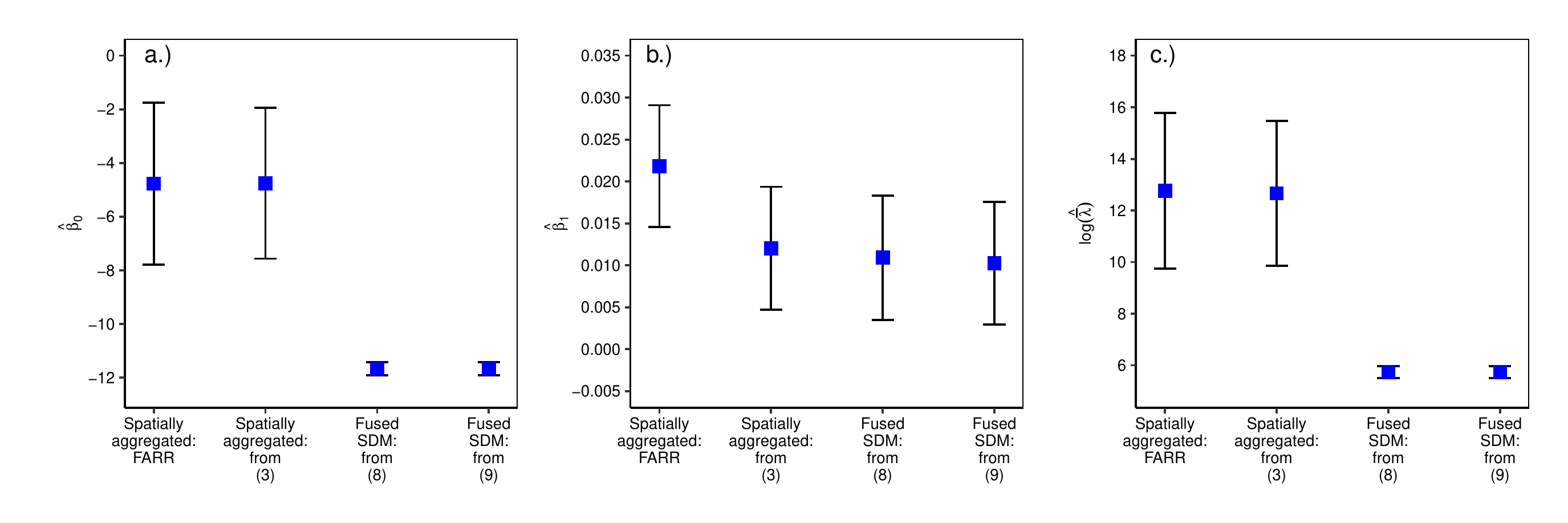}%
\renewcommand\thefigure{4} 
\caption{Panel (a), panel (b), and panel (c) display the parameter estimates and the 95\% CIs for the intercept ($\beta_0$), the relationship between the abundance and elevation ($\beta_1$), and the expected abundance ($\bar{\lambda}$) for Grasshopper Sparrows at Konza Prairie Biological Station, Kansas. The parameter estimates are obtained from the model proposed by \cite{farr2020integrating} for spatially aggregated data (Spatially aggregated: FARR), the model from (3) transformed for spatially aggregated data using change of support (Spatially aggregated: from (3)), our proposed fused data SDM from (8) without incorporating recorded distances (Fused SDM: from (8)), and our proposed fused data SDM from (9) incorporating recorded distances (Fused SDM: from (9)). The parameter estimates are shown by the blue square (\crule[blue]{2mm}{2mm}), and the 95\% CIs are shown by whisker ends.}
\label{fig:fig1:}
\end{sidewaysfigure}

\renewcommand{\arraystretch}{1}
\begin{table}[H]
\centering
\renewcommand\thetable{2:} 
\caption{Parameter estimates and the width of the 95\% CIs  for the intercept ($\beta_0$), the relationship between the abundance and elevation ($\beta_1$), and the log of the expected abundance ($\bar{\lambda}$) for Grasshopper Sparrows at Konza Prairie Biological Station, Kansas. The parameter estimates are obtained from the model proposed by \cite{farr2020integrating} for spatially aggregated data (Spatially aggregated: FARR), the model from (3) transformed for spatially aggregated data using change of support (Spatially aggregated: from (3)), our proposed fused data SDM from (8) without incorporating recorded distances (Fused SDM: from (8)), and our proposed fused data SDM from (9) incorporating recorded distances (Fused SDM: from (9)).
}
\npdecimalsign{.}
\nprounddigits{2}

\begin{tabular}{p{0.16\textwidth}|>
{\centering}p{0.11\textwidth}>
{\centering}p{0.09\textwidth}|>
{\centering}p{0.09\textwidth}>
{\centering}p{0.09\textwidth}|>
{\centering}p{0.09\textwidth}>
{\centering\arraybackslash}p{0.09\textwidth}}
\hline
\multirow{2}{*}{Models}&\multicolumn{2}{c}{$\beta_0$}&\multicolumn{2}{c}{$\beta_1$}&\multicolumn{2}{c}{$log(\bar\lambda)$}\\\cline{2-7}
& $\hat\beta_0$ &  Width of 95\% CI & $\hat\beta_1$ & Width of 95\% CI & $log(\hat{\bar\lambda})$ &  Width of 95\% CI\\
\hline
   Spatially 
   
   aggregated: 
   
   FARR & -4.767 & 6.034& 0.022 &   0.015& 12.766&  6.033 \\
  \hline
  Spatially 
  
  aggregated: 
   
   from (3) & -4.751& 5.616 & 0.012&  0.015& 12.669& 5.619\\
   \hline
     Fused SDM: 
    
    from (8) & -11.669 &  0.486 &0.011 &  0.015&  5.742 & 0.463\\
    \hline
   Fused SDM: 

from (9)&  -11.663 & 0.484 &0.010 &  0.015& 5.743 &  0.463 \\
\hline
\end{tabular}
\end{table}

\section{Discussion}

\subsection{IPPP generalization for DS and CR data that enables data fusion}

A critical aspect of data fusion is providing model representations for multiple data types that target the same inference. The existing point process based models for DS data use individual location information to infer about species-habitat relationship and abundance. In contrast, the existing point process based models for CR explicitly use home range centers. Therefore, the parameters in the underlying point process for the two data sources do not target the same inference. This incompatibility in the underlying process model may explain the lack of approaches for data fusion of DS and CR data. Our proposed approach provides a generalization of \cite{borchers2015unifying}'s IPPP based model with model representations for DS and CR data that share parameters in the underlying process that target the same inference, hence enabling data fusion. Therefore, our approach enables the use of these two types of high-quality planned survey data to obtain useful statistical inference regarding the species-habitat relationship, accurate estimates for the expected abundance, and more accurate spatial maps for species distributions.

\subsection{Improvement of inference regarding species-habitat relationship and estimate for the expected abundance by properly accounting for missing data issues}

Efficiently acquiring reliable parameter estimates for both $\beta_0$ and $\beta_1$ is of utmost importance. However, many recent studies only attempt to improve the estimate of $\beta_1$, focusing on species-habitat relationships or relative abundance (a measure of expected abundance relative to other species within a community). These approaches do not improve estimates of  $\beta_0$. In contrast to relative abundance, expected abundance plays a vital role in studying the dynamics of species populations, and estimating the expected abundance depends on both $\beta_0$ and $\beta_1$. It is also important to note that a small deviation of $\hat\beta_0$ and $\hat\beta_1$ from the true parameter value would significantly affect the estimate for the expected abundance due to the exponential function (i.e., $\hat{\lambda}(\mathbf s) = e^{\mathbf{x}(\mathbf {s})'\hat{\boldsymbol \beta}}$). Our study shows that obtaining reliable, more efficient parameter estimates for $\beta_0$ and $\beta_1$ crucially relies upon properly accounting for the missing data issues. Our modeling framework explicitly acknowledges and accounts for the missing data issues in DS and CR data using theory and tools from missing data literature.

Our results show that when the data contain partial location information, ad-hoc approaches such as spatial aggregation result in bias parameter estimates with poor efficiency (see Table 1). Our proposed models provide reliable, more efficient parameter estimates than existing approaches that use spatial aggregation (see Table 1). Furthermore, our simulation experiment led to an important finding: the inclusion of additional information regarding individual locations into the model, such as recorded distances, led to a significant efficiency gain in the parameter estimates. In fact, the efficiency gain surprisingly reaches the efficiency of the parameter estimates obtained under the benchmark scenario with complete location information.

\subsection{A spatio-temporal fused data SDM}

In our simulation experiment, the non-overlapping detection/capture regions ensure the independence of observations across and within surveys. In our data example, we assumed that the observations are independent across and within the surveys. However, we can strengthen the independence assumption by extending our model to a spatio-temporal model. A spatio-temporal model enables the modeling of species abundance patterns across both time and space. By using a continuous-space discrete-time model with short time periods, we can strengthen the independence assumption. However, a spatio-temporal model may have to address the spatio-temporal autocorrelation, which can be addressed by adding a spatio-temporal random effect. A bewildering number of approaches within the SDM literature are developed to model the spatial and spatio-temporal autocorrelation \citep[e.g.,][]{chakraborty2011point, renner2015point, mohankumar2021using}, which can be used to incorporate a spatial or a spatio-temporal random effect.

\subsection{Detection and capture functions}

In our study, we defined the probability of detection by a half-normal function of the distance between the point or the transect and the location of the individual. We defined the probability of capture as a constant parameter. However, the probability of detection can be defined by other functions such as uniform, hazard-rate, negative exponential, etc. Similarly, the probability of capture can be defined as a function of covariates such as the observer's experience level or environmental or geographical features. Such extensions of the model enable identifying the factors that influence the probability of detection or capture. 

It is possible that the parameters in the detection function or capture function are confounded with the parameters in the intensity function. For example, in a model in which the underlying intensity and the probability of capture are both functions of the same spatial covariate,  the underlying point process is confounded with the capture process. For another example, if the underlying intensity function is a function of the distance from the transect, the underlying point process is confounded with the detection process. Accounting for such confounding of the underlying intensity and the detection/capture probability is an area that needs further research. In most situations, we can avoid such confounding during the design of the surveys. 

\subsection{Inclusion of the spatial and non-spatial covariates}

The intensity function, probability of detection, and probability of capture can depend on many covariates that are spatial or non-spatial. For instance, in our Grasshopper sparrow data example, the practitioners may want to include "effort" to define the probability of detection, which is a non-spatial covariate, or they may want to include "vegetation," which is a spatial covariate. A non-spatial covariate that is measured during the survey can be easily incorporated into our model. However, for the spatial covariate, our approach requires the spatial covariate values for the entire study region. In most cases, they can be obtained from a geographical information system. However, obtaining the spatial covariate values in the entire study region can be trivial in some situations. In such situations, we can employ an auxiliary model to utilize the available data to predict the spatial covariate values for the entire region and use the predicted values as the input values for the spatial covariate in our models. 

\section*{Acknowledgements}

We thank all individuals, including R. Donnelly, J. Gresham, K. Kersten, A. Mayers, E. Smith, and M. Winnerman, who contributed to the Grasshopper Sparrows data used in our data example. This material is based upon work supported by the National Science Foundation under Grant No. 1754491 and the Konza Prairie Long-Term Ecological Research (LTER) Grant No. 1440484. This work was permitted by the US Geological Survey’s Bird Banding Lab (23836), the Kansas Department of Parks, Wildlife, and Tourism, and the Kansas State University Institutional Animal Care and Use Committee (IACUC) (protocol 4250).

\section*{Appendix A. Supplementary data}

Details associated with deriving our models are provided in the Supplementary Material. Annotated R codes, data, and additional files (e.g., shapefiles) that can be used to reproduce all results and figures associated with the simulation experiment, and the Grasshopper Sparrows data example are available in the Supplementary Material. 

\bibliographystyle{model5-names}
\bibliography{mybibfile}

\end{document}